# Size distributions reveal regime transition of lake systems under different dominant driving forces

**Shengjie Hu[1,2], Zhenlei Yang[2], Sergio Torres[2], Zipeng Wang[3], Ling Li[2]**

[1] College of Environmental and Resource Sciences, Zhejiang University.

[2] School of Engineering, Westlake University.

[3] School of Science, Westlake University.

Corresponding author: Ling Li (liling@westlake.edu.cn)

**Abstract**

Power law size distribution is found to associate with fractal, self-organized behaviors and patterns of complex systems. Such distribution also emerges from natural lakes, with potentially important links to the dynamics of lake systems. But the driving mechanism that generates and shapes this feature in lake systems remains unclear. Moreover, the power law itself was found inadequate for fully describing the size distribution of lakes, due to deviations at the two ends of size range. Based on observed and simulated lakes in China's 11 hydro-climatic zones, we established a conceptual model for lake systems, which covers the whole size range of lake size distribution and reveals the underlying driving mechanism. The full lake size distribution is composed of three components, with three phases featured by exponential, stretched-exponential and power law distribution. The three phases represent system states with successively increasing degrees of heterogeneity and orderliness, and more importantly, indicate the dominance of exogenic and endogenic forces, respectively. As the dominant driving force changes from endogenic to exogenic, a phase transition occurs with lake size distribution shifted from power law to stretched-exponential and further to exponential distribution. Apart from compressing the power law phase, exogenic force also increases its scaling exponent, driving the corresponding lake size power spectrum into the regime of "blue noise". During this process, the autocorrelation function of the lake system diverges with a possibility of going to infinity, indicating the loss of system resilience.

## 1 Introduction

Natural lakes are vital resources for human society, providing services such as water supply, food production and recreation (Downing et al., 2006; Feng et al., 2019). As an integral part of the hydrological and biogeochemical cycles, they are of growing interest with regard to their feedbacks to global climate change, particularly through carbon burial and emission (Rosentreter et al., 2021). An important basis for evaluating the ecological and economic values of lakes as well as their responses to climate change is to know how many lakes there are and how big they are – the lake size distribution (Verpoorter et al., 2014; Downing et al., 2006). Previous studies have shown that, like many complex systems, the size of lakes follows the power law distribution and this is an emergent feature shared by lake groups in different regions (Downing et al., 2006; Cael & Seekell, 2016; Steel & Heffernan, 2017; Feng et al., 2019). The power law size distribution of lakes provides an easy-to-use tool for estimating regional or global lake area and area-dependent variables, e.g., fluxes of greenhouse gases (GHGs), based on limited observations, thereby their roles in climate change (Seekell et al., 2014; Holgerson & Raymond, 2016). More importantly, the power law size distribution can be associated with crucial system behaviors including scale-invariance, critical transition and self-organization (Bak et al., 1987; Pascual et al., 2002; Rietkerk et al., 2021), and thus is thought to represent fundamental lake group characteristics that are essential for understanding the dynamics of lakes at the system level.

Although the power law size distribution holds promise both for attribute estimation and dynamics development of lake system, the realization of its potential has been hindered by the following two unsolved problems. One concerns the inadequacy of power law in describing lake size distribution over the whole size range, because of deviations at both lower (small lakes) and upper (big lakes) ends (Seekell & Pace, 2011; Mosquera et al., 2017). Given the importance of small and big lakes in the water and carbon cycles (Holgerson & Raymond, 2016; Messager et al., 2016), the assessments of lake systems in connection to water storage capacity and carbon budget would be subjected to bias if only power law size distribution is applied without considering the deviations, which in turn lead to uncertainties in evaluating lakes' role in global climate change. For example, the abundance of small lakes in a lake system was found to be overestimated when it was extrapolated from the power law size distribution determined by the sample set (Verpoorter et al., 2014), resulting in the overestimation of GHGs emissions from this system as a carbon source. Therefore, an accurate description of the full size distribution of lakes is necessary and essential.

The other problem with the lake size distribution is that it remains unknown what kind of driving factors and processes, and how they act and interact to produce the power law distributions observed in lake systems (Cael et al., 2022). The formation of lakes is premised on suitable topographic condition and sufficient water available, reflecting the joint control of endogenic and exogenic forces (Wirthmann, 2000). Hence, topography and precipitation should be two crucial factors for lake occurrence. In addition, human activities which are known to have locally altered the size and abundance of lakes also play a role (Steel & Heffernan, 2017). Processes shaping these factors are thus the primary driving processes of lakes. Although the size distribution of lakes is taken as the outcome of these driving factors and processes, the specific link between them has not yet been determined, because the isolation of the factors and processes is inherently difficult. As a result, limited insights into lakes' formation and development were obtained from the power law size distribution, even if its characteristics were observed to vary obviously with the location (McDonald et al., 2012; Feng et al., 2019). A mechanism that can explain the emergence of power law size distribution in lake systems and its regional variations is still missing in limnology.

In this study, we aimed to fill the aforementioned knowledge gaps in order to advance the understandings of lake size distribution as well as its driving mechanism. Through analyzing and characterizing the full size distributions of lakes observed in China's 11 zones with different hydrological and climatic conditions, we found that the lake size distributions are composed of three components over the whole size range and contain three phases featured by different probability distribution functions that represent distinct system properties. To uncover the driving mechanism underlying these statistical characteristics, potential lakes were simulated with consideration of only topographic suitability and compared with observed lakes spatially and statistically in each zone. Based on the comparison results along with precipitation data and satellite images, the driving factors, processes and forces of lake systems were quantitatively disentangled and linked with the characteristics of lake size distributions. To further integrate the results and findings from the analysis, we proposed a conceptual model that describes the full size distribution of lakes and shows how the size distributions could indicate and change with the system's dominant driving forces. By introducing a power spectrum analysis, we further explored the possible implications of the established model for studies of potential impacts of environmental changes on lake systems.

## 2 Study area and data

The study area covers China' 11 hydro-climatic zones that were divided based on the regional hydrology and climate (Xiong, 1995), with abundant natural lakes under diverse environmental and economic conditions (Feng et al., 2019). These zones were taken as the basic units for statistical and spatial analysis (Figure 1 and Table 1).

Data of observed lakes were taken from the widely used lake database, HydroLAKES, which provides global lake polygons and attributes with natural lakes and reservoirs included (Messager et al., 2016). The reliability of this global scale data within China has been validated by our previous study on the fractal scaling relationships of lakes (Hu et al., 2022). The natural lakes were extracted for each hydro-climatic zone and analyzed them as a lake system (Supporting Information (SI) Text S1). The reservoirs were extracted for the globe and China to test the hypothesis we proposed concerning impacts of human activities.

Digital elevation data, ASTER GDEM2, which covers the globe at the 30m resolution, was collected to simulate the distribution of potential lakes. A time series (1980-2015) of precipitation data across China (Wang et al., 2006) was used to quantify the contribution of rainfall to lake formation. As the resolution of this data is 1km, it was resampled to 30m to reconcile with the topography data prior to calculating the multi-year average precipitation (Text S2) for each lake and each hydro-climatic zone. The gross domestic product (GDP; Liu et al., 2005) data and human modification index (HMI) data were used to reflect the intensity of human influence in study zones (Table 1). The HMI data quantified the human modifications of land surface based on 13 anthropogenic stressors (Kennedy et al., 2019). To be comparable among regions, in each zone, the total amount of GDP was normalized by the zone area and the HMI was divided into three levels, i.e., "low" ($0 \leq HMI < 0.4$), "median" ($0.4 \leq HMI < 0.7$) and "high" ($0.7 \leq HMI \leq 1$) as adopted in the study of Kennedy et al. (2019) with proportion of areas under each category calculated (Table 1 and Table S1). Since both the GDP and HMI data were only used at the zone level (i.e., averaged across each zone), there is no need to perform resampling for the original resolution of 1km. To determine the lake-river connectivity, China's first-order stream data (Figure 1) was overlaid spatially with the observed lakes. All data processing and spatial analyses were performed in ArcGIS 10.2.

## 3 Methodology

### 3.1 Lake size distribution: probability distribution functions and fitting methods

The size distributions of observed lakes in China's 11 hydro-climatic zones displayed on log-log plots exhibit a segmented behavior over the whole size range, with the power law trend in the middle and deviations at the two ends (Figure 2). Thus, the full size distribution of lakes was determined for each zone in the following steps: (1) fitting power law distribution to the lake sample set and identifying the applicable size range of power law; (2) extracting lakes in the deviation ranges (lower and upper ranges separately) as independent subsets and analyzing their statistical distributions.

### 3.1.1 Power law distribution

Mathematically, a quantity $x$ conforms to power law distribution if its probability density function (PDF) is as Eq. (1) (Clauset et al., 2009). The power law size distribution of a system is usually described and displayed by the complementary cumulative distribution function (CCDF, Eq. (2)), because CCDF is more robust against the fluctuation caused by finite system sizes than PDF (Newmann, 2005; Clauset et al., 2009).

$$p(x) = \frac{\alpha-1}{x_{min}} \left(\frac{x}{x_{min}}\right)^{-\alpha} \quad (x > 0; x_{min} > 0; \alpha > 1) \tag{1}$$

$$P(x \geq X) = \left(\frac{X}{x_{min}}\right)^{b} \quad (X \geq x_{min}) \tag{2a}$$

with

$$b = \alpha - 1 \tag{2b}$$

where $p(x)$ and $P(x)$ are, respectively, the PDF and CCDF; $x_{min}$ represents the lower limit of power law distribution; $\alpha$ and $b$, the exponents of the two power functions, are related.

As shown in Eqs. (1)-(2), the applicability of power law distribution to a sample set has a lower limit ($x_{min}$), that means, power law behavior occurs in a certain range ($x \geq x_{min}$) of the distribution. The linear regression approach commonly adopted in previous studies (Cael et al., 2017; Feng et al., 2019; Steele & Heffernan, 2019) was not used here, because it cannot determine the value of $x_{min}$ and is likely to give biased estimate of the power exponents ($\alpha$ and $b$) (Newmann, 2005; Clauset et al., 2009). Instead, a statistical framework that combines maximum-likelihood fitting methods, Kolmogorov-Smirnov statistics and likelihood ratios (Clauset et al., 2009) was applied to fit power law distribution (Eq. (1)) to the lakes in our study zones (Alstott et al., 2013). This method is able to directly output the values of $x_{min}$ and $\alpha$. $x_{min}$ divides the lake sample set into two subsets: the deviated subset below the lower limit ($x < x_{min}$) and the power law-distributed subset ($x \geq x_{min}$) with sample sizes denoted as $N_1$ and $N_2$, respectively. The conformity of lakes to the power law size distribution in a hydro-climatic zone then can be quantified by the conformity ratio, $R$, defined in Eq. (3).

$$R = \frac{N_2}{N_1 + N_2} \tag{3}$$

Deviations were observed at the upper end of power law size distributions in several hydro-climatic zones (Figure 2). It is known that the finite size effect could cause a downward trend in the tail of the power law size distribution of a system (Chen et al., 2021). However, downward, upward and no deviation were shown in our study sites (Figure 2 and Table 2), which cannot be fully explained by the finite size effect and need to be analyzed separately. Since no method can automatically identify the upper deviation point ($x_{max}$), we visually determined the value of $x_{max}$ on the lake size distribution plot, and then extracted the lakes with areas greater than $x_{max}$ for analysis of the size distribution. These lakes were found to also obey the power law distribution (Figure S1). Since the sample size should not be less than 100 for the statistical significance of power law fitting (Clauset et al., 2009), the size distribution of lakes in the upper deviation range was characterized and discussed only for the two Tibetan Plateau zones (Qiangtang Basin Zone, QB; and Eastern and Southwestern of Tibet Zone, EST), where the sample sizes nearly meet this requirement.

### 3.1.2 Exponential and stretched-exponential distribution

Lakes deviated from power law below the lower limit ($x_{min}$) constitute the subset with sample size of $N_1$. The portion of these lakes in the whole sample set of a study zone is given by the width ratio (WR, Eq. (4)) and their size distribution can be fitted either to the exponential or to the stretched-exponential distribution (Eqs. (5)-(6)). The fitting was conducted in Python 3 (https://www.python.org/) via the scipy.optimize.curve_fit model.

$$WR = 1 - R = \frac{N_1}{N_1 + N_2} \qquad (4)$$

$$P(x) = C_1 e^{-c_1 x} \qquad (5)$$

$$P(x) = C_2 x^{\beta-1} e^{-c_2 x^\beta} \qquad (6)$$

where $P(x)$ represents the CCDF; $C_1$, $c_1$, $C_2$, and $c_2$ are fitting parameters; and $\beta$ is the exponent of stretched-exponential distribution.

The goodness of fit of the exponential or stretched-exponential distribution was evaluated by coefficient of determination ($r^2$, Eq. (7)) and Root Mean Squared Error (RMSE, Eq. (8)).

$$r^2 = 1 - \frac{\Sigma_i(y_i - y'_i)^2}{\Sigma_i(y_i - \bar{y})^2}, \text{ with } \bar{y} = \frac{1}{N_1}(\Sigma_i^{N_1} y_i) \qquad (7)$$

$$RMSE = \sqrt{\frac{1}{N_1} \Sigma_i^{N_1}(y_i - y'_i)^2} \qquad (8)$$

where $y_i$ and $y'_i$ represent the observed and predicted value, respectively, and $\bar{y}$ is the arithmetic mean of the observed values. Since the variable examined here is the cumulative probability of the lake size, the RMSE has no dimension.

### 3.2 Power spectrum analysis

### 3.2.1 Power spectrum

If a quantity follows the power law distribution, its power spectrum will also be a power function (Sornette, 2006). Based on the transformation equation used by Bak et al. (1987), we computed the power spectrum ($S(\varphi)$) of lake size, as follows,

$$S(\varphi) = \int dx \frac{xp(x)}{1+(\varphi x)^2} \propto \varphi^\gamma \qquad (9a)$$

with

$$\gamma = -2 + \alpha \qquad (9b)$$

where $p(x)$ is the power law distribution of lake size, $\varphi$ represents the transformed quantity corresponding to $x$ (lake size) and $\gamma$ is the exponent of lake size power spectrum.

The type of power spectrum can be determined by the value of its exponent ($\gamma$) (Halley, 1996). If $\gamma = 0$, the power spectrum is a white noise type; if $-1 \leq \gamma < 0$, it belongs to the pink noise; and if $\gamma > 0$, it is classified as the blue noise. More information about the definitions of the different types of power spectrum can be found in SI Text S3.

### 3.2.2 Hurst exponent

For a power spectrum in the fractional Gaussian noise category (i.e., with $-1 < \gamma < 1$), its Hurst exponent (H) can be calculated through the equation below (Womell, 1993)

$$H = \frac{\gamma - 1}{2} + 1 \qquad (10)$$

According to the fractional Brownian motion theory, $H = 0.5$ indicates that there is no correlation between processes, $H > 0.5$ indicates that the correlation between processes is persistent, and $H < 0.5$ indicates that the correlation is anti-persistent (Mandelbrot & Van Ness, 1968).

### 3.2.3 Autocorrelation function

Autocorrelation function provides a way for detecting critical transition of complex systems (Scheffer et al., 2009). The increase of autocorrelation typically indicates the loss of the system's resilience (Text S1) and a potential tipping for the system to an alternative state (Scheffer et al., 2009). Autocorrelation function ($R(\tau)$) and power spectrum ($S(\varphi)$) are a Fourier pair as described by the Wiener-Khinchin equation (Eq. (11); Kubo & Hashitsume, 2012). This relationship can be described as Eq. (12), if $R(\tau)$ and $S(\varphi)$ are even functions, according to the Euler formula.

$$R(\tau) = \frac{1}{2\pi} \int_{-\infty}^{+\infty} S(\varphi) e^{i\varphi\tau} d\varphi \tag{11}$$

$$R(\tau) = \frac{1}{\pi} \int_{0}^{+\infty} S(\varphi) \cos \varphi\tau \, d\varphi \tag{12}$$

With the assumption of even functions, the analytical expression of the autocorrelation function corresponding to the lake size power spectrum was derived (Eq. (13a)) and its behaviors under different types of power spectrum were analyzed. It was found that when $S(\varphi)$ is a white noise type ($\gamma = 0$), $R(\tau)$ is a $\delta$ function (Eq. (13b)); when $S(\varphi)$ belongs to pink noise ($\gamma < 0$), $R(\tau)$ goes asymptotically to zero (Eq. (13c)); and when $S(\varphi)$ belongs to blue noise ($\gamma > 0$), $R(\tau)$ diverges, with a possibility of approaching infinity (Eq. (13d)).

$$R_x(\tau) = \frac{1}{\pi} \int_0^{+\infty} \varphi^\gamma \cos \varphi\tau \, d\varphi$$

$$= \frac{1}{\pi\tau} \varphi^\gamma \sin(\varphi\tau)|_{\varphi \to +\infty} + \frac{\gamma}{\pi\tau^2} \varphi^{\gamma-1} \cos(\varphi\tau)|_{\varphi \to +\infty} + \frac{\gamma(\gamma-1)}{\pi\tau^3} \varphi^{\gamma-2} \sin(\varphi\tau)|_{\varphi \to +\infty}$$

$$+ \frac{\gamma(\gamma-1)(\gamma-2)}{\pi\tau^3} \int_0^{+\infty} \sin(\varphi\tau) \varphi^{\gamma-3} d\varphi \quad (-1 < \gamma < 1) \tag{13a}$$

$$\text{For } \gamma = 0, \ R(\tau) = \frac{1}{\pi\tau} \sin(\varphi\tau)|_{\varphi \to +\infty} = \delta(\tau) \tag{13b}$$

$$\text{For } \gamma < 0, \ R(\tau)|_{\varphi \to +\infty} \to 0 \tag{13c}$$

$$\text{For } \gamma > 0, \ R(\tau) \approx \frac{1}{\pi\tau} \varphi^\gamma \sin(\varphi\tau)|_{\varphi \to +\infty} \ (diverge) \tag{13d}$$

### 3.3 Phase definition and properties

The three probability distributions found in the whole size range of lake size distribution represent different phases of lake (size) state (definitions of phase and state are in SI Text S1). The properties of these phases were examined through analyses of fluctuation and entropy of the size distribution, which are commonly used to describe the system state in statistical physics. Fluctuation was assessed based on coefficient of variance (CV, Eq. (14)), which is the ratio of the standard deviation ($\sigma$) to the mean ($\mu$), measuring the dispersion of a probability distribution (Loreau & Mazancourt, 2013). Entropy quantifies the chaos degree of a system (Lesne, 2014). Average entropy (AE, Eq. (15)) was calculated by normalizing the Shannon entropy of the lake size distribution using the sample size to eliminate the difference caused by the lake abundance disparity among regions.

$$CV = \frac{\sigma}{\mu} \tag{14}$$

$$AE = \frac{1}{N}[-\sum_{i=1} n_i p_i(x) \log_2 p_i(x)] \tag{15}$$

where $\sigma$ and $\mu$ are the standard deviation and the mean of lake size in a sample set; $N$ (= $N_1$ + $N_2$) is the sample size; $n_i$ is the lake abundance corresponding to the $i$-th lake size; and $p_i(x)$ represents the probability of the $i$-th lake size in the sample set.

### 3.4 Identification and isolation of driving factors

To unveil the mechanism underlying the statistical characteristics and properties of lake size distributions, we need to identify the driving factors of lake systems and quantify their controls on lakes. A procedure was applied to first examine the effect of topography followed by precipitation and other unknown factors (Figure 3). The analysis on the effects of topography and precipitation was based on the simulation of potential lakes, and comparison of average precipitation between the lake and the zone, respectively, while the identification and quantification of other factors were realized with the help of satellite images (i.e., Landsat images). The proportion of lakes under the influence of each factor in a sample set represents the control of the factor on the lake system within the study zone.

#### 3.4.1 Potential lake simulation and topographic control

Potential lakes are defined as the lakes that would form as long as the topographic condition is suitable (Hu et al., 2017). They were simulated via the topographic wetness index (TWI, Eq. (16)), which reflects the control of topography on hydrological processes (Beven & Kirkby, 1979).

$$TWI = log_{10}(\frac{AS}{tan\theta}) \tag{16}$$

where $AS$ is the accumulation area of upstream pixels with unit of m$^2$ and $tan\theta$ is the local topographic slope. In the simulation, $AS$ was calculated using the D-8 algorithm, in which the water of a pixel is transferred along the steepest downslope direction to one of the eight neighboring pixels. Thus, the value of $AS$ is equal to the product of the number of upstream pixels that can accumulate into a pixel and the pixel area (e.g., since the resolution of the DEM data used is 30m, the area represented by each pixel is 30×30m$^2$).

The value of TWI indicates the wetness of a location, namely, the greater the value, the wetter the location will be (Beven & Kirkby, 1979). The key step in simulating potential water bodies using TWI is to determine an appropriate threshold that can distinguish the target type from other land forms/covers (Hu et al., 2017). By analyzing the frequency distribution of TWI values (Figure S2) and cross-validation with Landsat images, we found that TWI = 12.71 can well distinguish lakes from other land covers in all the hydro-climatic zones (Figure S3), and hence was set as the threshold. The sample set of simulated potential lakes for each zone was established by extracting the pixels with TWI values greater than this threshold while keeping the minimum lake area consistent with the observed ones in order to avoid the inconsistency caused by data resolution differences (Figure S4). In addition, considering rivers and reservoirs may be included in the simulation results, we manually checked and removed them based on the reservoirs contained in the HydroLAKES and the Landsat images. The size distribution of potential lakes was analyzed using the same method as for the observed ones.

#### 3.4.2 Isolation of different driving factors

Since potential lakes were simulated based on topographic suitability only, they are supposed to be representative of topography. Thus, observed lakes that spatially overlap with simulated potential lakes are considered to be topography-controlled (Figure 3). The proportion of these overlapped lakes quantifies the contribution of topography to the lake formation in the zone, while the non-overlapped ones are attributed to other factors besides topography. As sufficient water supply is also an important factor in lake formation, the influence of precipitation was examined subsequently. Using the 36-year precipitation dataset of China (Wang et al., 2006), the average precipitation was calculated for each lake and each zone (Figure 3). A non-overlapped lake is considered to be in the control of precipitation in two cases: (1) the average precipitation at the lake greater than that of the whole hydro-climatic zone explains why a lake was observed but not simulated; and (2) the average precipitation at the lake lower than that of the whole hydro-climatic zone explains why a lake was simulated but not observed. For other non-overlapped lakes that could not be explained by the precipitation criteria, the influence of other factors was considered. By overlaying these lakes with high resolution satellite images, we visually interpreted the factors contributing to the mismatch. All these factors can be summarized into seven categories, including human activities, other water supply, unclosed topography, interannual fluctuation, type conversion, data uncertainty and special cases. More details about these seven other factors are provided in SI (Text S4 and Figure S5).

## 4 Results and discussion

### 4.1 Full size distribution of lakes: components and phases

The statistics of the observed lakes in China's 11 hydro-climatic zones showed that the lake size distributions can be decomposed into three components over the whole size range, described here as shoulder ($x < x_{min}$), body ($x_{min} \leq x \leq x_{max}$) and tail ($x > x_{max}$), which obey distinct probability distribution functions (Figure 4). The shoulder fits either the exponential or stretched-exponential distribution, while the body and tail both follow the power law distribution (Figure 4, Figure S6 and Figure S7). We suggest that these three statistical distributions represent three different phases of lake size distribution. The fitting of power law distribution to the whole lake sample set using the method of Clauset et al. (2009) as described in section 3.1.1 determined $x_{min}$ that separated the shoulder from the power law-distributed components.

The fitting of the exponential distribution or stretched-exponential distribution to the shoulders is related to the width ratio (*WR*), which is the portion of shoulder lakes in the whole lake sample set. The shoulders of hydro-climatic zones with high *WR* ($\geq 60\%$) fit well to the exponential distribution, and elsewhere ($WR < 60\%$) the stretched-exponential distribution holds (Table 2), with $r^2$ greater than 0.98 and RMSE less than 0.01 for the fitting in all cases (Figure S6). Since shoulder lakes and power law-distributed lakes constitute the entire sample set (Eq. (4)), this result also indicates that in hydro-climatic zones with less power law-distributed lakes, shoulder lakes fit the exponential distribution.

The power law-distributed lakes reside in both the body and tail components, but mainly in the former with a large proportion over 90%. Thus, the power law behavior of lake size distribution is characterized by two parameters of the subset of $x \geq x_{min}$ – the conformity ratio $R$ and the scaling exponent $b_1$, both of which were found to vary with the location. Great disparity in $R$ was shown among the study zones, from 7.25% for North China Zone (NC) to 91.36% for QB (Figure 1 and Table 2). The scaling exponent partitions the 11 hydro-climatic zones into two classes: C1 with $b_1$ less than 1 and C2 with $b_1$ larger than 1 (Table 2). To further distinguish C1 and C2, we calculated the lake size power spectrum and the corresponding Hurst exponent (H) for each zone. The results uncovered contrasting features of observed lakes in these two classes (Figure 5). In C1, the exponents of lake size power spectrums are negative, classified as pink noise; and the corresponding H are less than 0.5, indicating an anti-persistent coupling of driving processes. In C2, the size power spectrums have positive exponents and are blue noise alike; and the corresponding H are larger than 0.5, an indication of persistent coupling of driving processes.

Deviation from the body at the upper end ($x_{max}$) was observed in half of the hydro-climatic zones, which cannot be fully explained by the finite size effect because of the inconsistent trends (Figure 2 and Table 2). Lakes in the tails ($x > x_{max}$) also conform to the power law distribution and show 100% conformity ratio (Figure 4b and Figure S1). Despite that the scaling exponents of tails ($b_2$) are different from those of their corresponding bodies ($b_1$), the tail lakes account for less than 10% of the number of lakes in the range of $x \geq x_{min}$ (Text S5) and hence affect little the characterization of the power law behavior of lake size distribution, which was performed to the whole power law-distributed subset (Figure S7).

The three probability distributions found in the size distribution are considered to be the three phases of lake (size) state. To further characterize these phases, coefficient of variation (CV) and average entropy (AE) were computed based on the observed lake data. The values of these two parameters indicate the degrees of system heterogeneity and orderliness, respectively (Loreau & Mazancourt, 2013; Lesne, 2014). The results showed the largest CV for the power law phase and the smallest for the stretched-exponent phase (Table 3). This is consistent with the previous study of Mazzarisi et al. (2021), which suggested that the power law distribution describes ensembles of the maximum diversity. The minimum AE is found in the power law phase and the maximum AE in the stretched-exponential phase (Table 3). This also agrees well with previous findings that the power law distribution arises from patterned, self-organized systems (Villegas & Caldarelli, 2021), whereas the stretched-exponential distribution is typically associated with disordered systems (Zaccone, 2020) and the exponential distribution often appears in stochastic events (Cox, 1995). According to the thermodynamics, as a system approaches the equilibrium (Text S1), the entropy will be maximized while the fluctuation decreases (Kleidon, 2010). Thus, we infer that the power law size distribution represents a heterogeneous, ordered phase that is far away from the equilibrium (Text S1), while the stretched-exponential phase is relatively homogenous, disordered and near the equilibrium, and the exponential phase may reflect an intermediate, random state.

**4.2 Driving mechanism underlying lake size distribution**

**4.2.1 Driving factors, processes and forces**

By simulating potential lakes and comparing them with the observed ones along with precipitation data and Landsat images, we identified and examined the driving factors of lake systems including topography, precipitation and seven other factors. The controls of these factors on lakes in a hydro-climatic zone were indicated by the proportion of lakes affected by each factor in the sample set. Topography is found to be the dominant factor in lake formation in all zones with an average contribution of 82%; precipitation and other factors account for 9% and 9%, respectively (Table S2). In general, endogenic and exogenic forces (Text S1) jointly shape most of the Earth's landforms (Wirthmann, 2000). Both of them can alter the topography; however, in light of the former's leading role over a variety of scales, topography and its shaping processes (topographic processes) are considered to represent mainly endogenic force here, whilst precipitation and the other factors together with their shaping processes (non-topographic processes) constitute the exogenic force regime (Figure 3). Therefore, the lake system as a whole is under the joint control of endogenic and exogenic forces, but the former plays a major role.

### 4.2.2 Dominant driving force of different phases

For the different phases of lake size distribution, our results revealed the dominance of different driving forces. Specifically, we found that the power law phases in all the hydro-climatic zones are mainly controlled by endogenic force, while the role of exogenic force is more profound in the other two phases (Figure 6 and Table 3). On average, 94.19% of the lakes in the power law phase are dominated by endogenic force. A similarly high percentage of lakes (94.07%) in the exponential phase are controlled by exogenic force. In contrast, the stretched-exponential phase appears to be more balanced, with 58.06% and 41.94% of lakes affected by exogenic and endogenic force, respectively. Combining these results with the phase properties, we argue that endogenic force tends to produce the ordered state of lake system as represented by the power law distribution, while exogenic force leads to the stochastic state following the exponential distribution, and the combined influence of these two forces is probably responsible for the disordered, stretched-exponentially distributed state.

To further verify this hypothesis, we analyzed the size distribution of reservoirs from the same dataset as observed lakes. On both global and national (China) scales, the reservoirs were found to conform to the stretched-exponential distribution consistently (Figure S8). This result provides supporting evidence for the hypothesis; because the reservoirs are built by humans on the premise of suitable topographic condition, they are in the transition zone jointly influenced by endogenic and exogenic forces and show the stretched-exponential size distribution. Our hypothesis is also in line with the geomorphology antagonism principle proposed by Scheidegger (1986), which stated the systematic nature of endogenic force and randomness associated with exogenic force.

### 4.2.3 Phase transition occurring with dominant driving force change

The conformity ratio ($R$) to power law correlates with the position of phase boundary ($x_{min}$) negatively (Figure S9), implicating that the increase of $x_{min}$ will contract the power law phase but expand the other two phases residing on the shoulder component. Compared with the values of simulated potential lakes, $x_{min}$ of observed lakes are larger in most hydro-climatic zones (Figure 5c), suggesting that exogenic force (absent in the simulation but existing in reality) drives lakes out of the power law phase and into the shoulder. The three exceptions (Southeast of Yunnan and Tibet Zone, SEYT; Northwest Mountainous Zone, NWM; Northwest Basin Zone, NWB) are possibly caused by the complex coupling between endogenic and exogenic forces in shaping the topography. We also noticed that hydro-climatic zones with large $x_{min}$ (>0.30 km$^2$) and low $R$ (<40% equivalent to $WR \geq 60\%$) all locate in China's developed regions, where the levels of GDP and human modification are high (Figure 1, Table 1 and Table S1). The shoulders of these regions all fitted to the exogenic force-dominated exponential size distribution (Table 2 and Figure S6). In contrast, the corresponding potential lakes conform better to the stretched-exponential distribution with larger value of r$^2$ (Table 2 and Figure S6). These results together demonstrate that the change of dominant driving force of lake system from endogenic to exogenic is likely to induce a phase transition (Text S1) with size distribution shifted from power law to stretched exponential and further to exponential distribution.

### 4.2.4 Characteristics of power law phase

Overall, the scaling exponents ($b_1$) of the power law phase of simulated potential lakes are smaller than their observed counterparts and less than 1 in most hydro-climatic zones, i.e., with "pink" power spectrum and anti-persistent coupling of driving processes (Figure 5 and Figure S7). Since potential lakes are driven by topographic processes, we could infer that topographic processes (endogenic force and factors) tend to generate "pink" size power spectrums and operate in an anti-persistent coupling mode, whereas non-topographic processes (exogenic force and factors) result in an opposite tendency with increased $b_1$ values. Therefore, the value of scaling exponent may indicate the influence of different types of driving processes on the power law phase.

Specifically, observed lakes in hydro-climatic zones belonging to C1 were found to maintain their potential counterparts' "pink" size spectrums and anti-persistent coupling features, but the increased scaling exponents make the spectrums less "pink" (Figure 5). Although the trend in C2 is not as clear, a "blue shift" (i.e., the size power spectrum turns from pink to blue noise or the blue size power spectrum becomes "bluer" with increased value of power exponent $\gamma$) in the power spectrum appears to occur to the observed lakes compared with their potential counterparts (Figure 5). The size spectrums in NC and Northeast Zone (NE) turn from "pink" (anti-persistent coupling) in potential lakes into "blue" (persistent coupling) in reality. The dramatic (color) change of the size spectrum may be linked to strong influence of non-topographic processes associated with human interference, particularly in well-developed NC (Table 1, Table S1 and Table S2). Surprisingly, the Qinba - Dabie Zone (QD) as the most developed region in China (Table 1 and Table S1, Zhou et al., 2018) retains consistency between its observed and potential lakes. The difference between QD and NC may be explained by their distinct lake-river connectivity (Table 1). The proportion of lakes connected with the first-order streams in QD is the highest in China; the ratio is about 8.7% for the power law-distributed lakes and 5.3% for the whole lake samples. These ratios are about 5.26% and 2.58% in NC, respectively. The only exception in the "blue shift" trend is SEYT, which may be explained by the extremely complex topography in this region (Xing & Ree, 2017), having the smallest area but the highest topographic relief (over 300m, Figure 1). From these results we can see that although the power law phase is dominated by endogenic force, the influence of exogenic force can be detected through the value of scaling exponent. Topographic processes tend to generate power law size distributions with scaling exponent less than 1 and keep the corresponding size power spectrums in the regime of pink noise; whereas non-topographic processes are more likely to increase the scaling exponent, shifting the size power spectrum from "pink" to "blue".

The features of tail component show little difference between observed and potential lakes in the same hydro-climatic zone with similar $x_{max}$ values and tail directions, but vary among different zones (Figure S7). The analysis of the two Tibetan plateau zones (EST and QB, Figure 4b, Figure S1 and Table 2) revealed quite different tail features even though these two zones are geographically adjacent. The $x_{max}$ values in QB and EST differ by an order of magnitude and their tail directions are opposite, which result in the opposite trend of changes in their scaling exponents and size power spectrums compared with those of the bodies (Figure 4b and Table 2). In QB, the scaling exponent of the tail increased and the size power spectrum shifted to the blue noise ($b_2 > 1 > b_1$), whereas in EST, the scaling exponent of the tail decreased and the size power spectrum became more "pink" ($b_2 < b_1 < 1$). The finite size effect (Chen et al., 2021) alone cannot fully explain these phenomena. Given that the tail component is composed of big lakes which are largely controlled by endogenic force and resistant to exogenic force, and different tectonic movements have been reported in QB and EST (Liu & You, 2015), we suggest that tectonic activities in the region may also influence the tail features. However, more research is needed to draw conclusions.

### 4.2.5 Conceptual model

Based on the above results and discussion, we proposed a conceptual model to describe the full size distribution of lakes and the underlying driving mechanism (Figure 7). Endogenic and exogenic forces were found to jointly shape the lake size distribution, but play different roles in its different phases. In the regime of endogenic force, lakes conform to power law size distribution, which is ordered and far away from the thermodynamic equilibrium. The scaling exponent ($b_1$) of power law phase marks the influence from topographic and non-topographic processes. If $b_1$ is less than 1, topographic processes are dominating; if $b_1$ is larger than 1, the influence of non-topographic processes intensifies. Two power law distributions with different scaling exponents may coexist in this phase, separated by a critical scale ($x_{max}$) which is likely to be affected by both the tectonic activities and the finite size effect of the region. As the influence of exogenic forces increases, the lake size distribution is likely to shift from power law to stretched-exponential and further to exponential phase. When the control of endogenic force is balanced by that of exogenic force, the stretched exponential phase appears, with properties of homogeneity and disorder. When exogenic force becomes dominant, the size distribution enters the exponential phase, which is considered to represent a random lake state. Overall, exogenic force tends to shrink the power law phase by driving lakes into the exponential phase.

### 4.3 Implications and limitations

Small lakes, especially those smaller than 0.001 km2, usually have a strong GHGs emission capacity, due to their fast hydrological and biogeochemical cycling rates (Holgerson & Raymond, 2016). However, because they are difficult to map, their areas and area-dependent variables (e.g., GHGs emissions) are often extrapolated from observed lake data (Rosentreter et al., 2021; Holgerson & Raymond, 2016; Seekell et al., 2014). It has been found that using the power law size distribution obtained by fitting the entire sample set to estimate these quantities for small lakes can be highly biased, as deviations from the power law are widely observed for small lakes (Seekell et al., 2014). Thus, there is a need for a better understanding of the size distribution of small lakes in order to reduce uncertainty in the estimation of GHGs emission from inland waters (Rosentreter et al., 2021; Holgerson & Raymond, 2016; Seekell et al., 2014). The full size distribution of lakes found here suggests that for small lakes the exponential or stretched-exponential distribution is a better statistical model for describing their sizes, providing a useful tool for further estimating lakes' role in climate change.

Apart from shrinking the power law phase, exogenic force also increases its scaling exponent. The shrinkage of power law phase corresponds to the decrease of ordered configuration and the increasing scaling exponent drives the lake size power spectrum into the regime of blue noise. Based on the Fourier transform, we found that when the size power spectrum is "pink", the autocorrelation function of lake system approaches zero asymptotically (Eq. (13c)); however, when shifted to "blue", the autocorrelation function diverges with a possibility of going to infinity (Eq. (13d)). According to the critical tipping theory, when a system loses its resilience (i.e., the ability of the system to maintain its current state or functions in the face of disturbances, Text S1), it exhibits critical slowing down, i.e., a prolonged recovery of the system from the disturbance and with an increase in autocorrelation (Scheffer et al., 2009).

Based on these results, we infer that ongoing and future environmental changes, which are in the regime of exogenic force, may make the lake systems less resilient.

The power law fitting method adopted here is designed to optimize the estimation accuracy of power exponent, and thus may result in a relatively large value for $x_{min}$ (Clauset et al., 2009). This will cause uncertainties in the calculation of the conformity ratio ($R$) of power law distribution. Moreover, there is no method that could automatically determine the upper limit ($x_{max}$) of power law size distribution, restricting the accurate separation and characterization of the two co-existing components in the power law phase. These limitations can be solved in two ways: one is to develop a better power law fitting approach and the other one is to find out a continuous function that can unify the components of the lake size distribution across the whole range.

The conceptual model presented here aims to establish links between the driving forces and lake size distribution. The applicability of the model was preliminarily explored here through predicting the feedback between lake system and environmental changes; however, it should be noted that this conceptual model is not yet able to make quantitative predictions. Building a dynamical model with prediction capacity for complex systems is still one of the grand challenges in the scientific community (Borner, 2021).

## 5 Conclusions

By studying observed and simulated lakes in China's 11 hydro-climatic zones statistically and spatially, we revealed the driving mechanism underlying the size distribution of lakes and built a conceptual model to link the characteristics of lake size distribution to the system's driving forces. The main findings are summarized as follows:

(1) Over the whole size range, the lake size distribution is constituted by three components, including shoulder (small range), body (medium range) and tail (large range). The shoulder follows either the exponential or the stretched-exponential distribution, whereas the body and tail both conform to the power law distribution. Moreover, hydro-climatic zones with less power law-distributed lakes tend to have shoulders that fit the exponential distribution.
(2) The three probability distributions contained in lake size distribution are the three phases that represent lake (size) state with different degrees of heterogeneity and orderliness. Specifically, the power law phase is heterogeneous, ordered and far away from equilibrium, while the stretched-exponential phase is relatively homogenous, disordered and near the equilibrium, and the exponential phase represents a medium, random state.
(3) The lake systems in the study zones are affected by topography, precipitation and other seven factors with average contribution of 82%, 9% and 9%, respectively, reflecting the joint control of endogenic and exogenic force on natural lakes.
(4) Endogenic force dominates the power law phase, while the influence of exogenic force is more significant in the other two phases, especially in the exponential phase. When the dominant driving force of lake system changes from endogenic to exogenic, the size distribution shifts from power law to stretched-exponential and finally to exponential phase.
(5) The scaling exponent of power law phase and the type of size power spectrum indicate the action of different types of driving processes. Topographic processes tend to produce power law phases with scaling exponents less than 1 and maintain size power spectrums in the regime of pink noise. In contrast, non-topographic processes increase the value of scaling exponent and drive size power spectrum into the regime of blue noise.

(6) Environmental changes, such as global warming and human interference, are likely to reduce the resilience of lake systems.

## Acknowledgments

The authors acknowledge funding support from National Natural Science Foundation of China (Grant No.41976162).

## Open Research

All data are publicly available. The lake dataset, HydroLAKES, was downloaded from HydroSHEDS (https://www.hydrosheds.org/page/hydrolakes). ASTER GDEM2 was produced and distributed by NASA (https://search.earthdata.nasa.gov/search/granules?p=C1575733858-LPDAAC_ECS&pg[0][v]=f&pg[0][gsk]=-start_date&q=C1575733858-LPDAAC_ECS&tl=1640038601.818!3!!). Human modification index was from the Global Human Modification of Terrestrial Systems Dataset, which was distributed by the Socioeconomic Data and Application Center (https://sedac.ciesin.columbia.edu/data/set/lulc-human-modification-terrestrial-systems). China's grid GDP data (https://www.resdc.cn/DOI/DOI.aspx?DOIID=33) and first-order river distributed data (https://www.resdc.cn/DOI/DOI.aspx?DOIID=44) are provided by the Resource and Environment Science and Data Center of Institue of Geographic Science and Natural Resources Research, Chinese Academy of Sciences. The potential lakes simulated in this study are available at https://doi.org/10.6084/m9.figshare.20224434.v1. The python package used to fit power law distribution to lake size data can be accessed from GitHub (https://github.com/jeffalstott/powerlaw).

**References from the Supporting Information**

**Figure. 1.** Spatial distribution of observed (natural) lakes in China's eleven hydro-climatic zones. The red dot represents the GDP of each zone and the size of the dot corresponds to the amount of GDP. The larger the dot, the richer the region is. The conformity ratio of power law phase ($R$) is given in the bracket.

**Figure. 2.** Size distributions of observed lakes in the hydro-climatic zones (the red line indicates power law distribution).

**Figure. 3.** Procedure for isolating driving factors, processes and forces of lakes.

**Figure. 4.** Schematic diagram of lake size distribution along with data analysis examples. (a) the three components (shoulder, body and tail) and three phases (I: exponential; II: stretched-exponential; III: power law); (b) Data fitting results from four hydro-climatic zones. ① Exponential phase exemplified by the lakes in the shoulder component of the Southeast Zone (SE) zone; ②Stretched-exponential phase exemplified by the lakes in the shoulder component of the Northwest mountainous Zone (NWM); ③ - ⑥ : Comparison of the two power law distributions coexisting in the power law phase between the two Tibetan Plateau zones (Qiangtang Basin Zone, QB; and East and southwest of Tibet Zone, EST).

**Figure. 5.** Power spectrum analysis results of observed and potential lakes. (a) Definition of colored noise; (b) and (c): Comparison of size power spectrum characteristics between observed and potential lakes among hydro-climatic zones.

**Figure. 6.** Size distribution phases (a), its properties (b) and dominant driving forces (c). The CV and AE values in (b) are the average of the 11 hydro-climatic zones, wherein the value of AE was multiplied by 100 to be better displayed.

**Figure. 7.** Conceptual model of the driving mechanism underlying lake size distribution.

**Table 1.** Eleven hydro-climatic zones of China.

**Table 2.** Characteristics of the three components and phases of lake size distribution.

**Table 3.** Properties (coefficient of variation (CV) and average entropy (AE)) and dominant driving force of different phases in each hydro-climatic zone. The value of AE was multiplied by 100 to be better displayed.

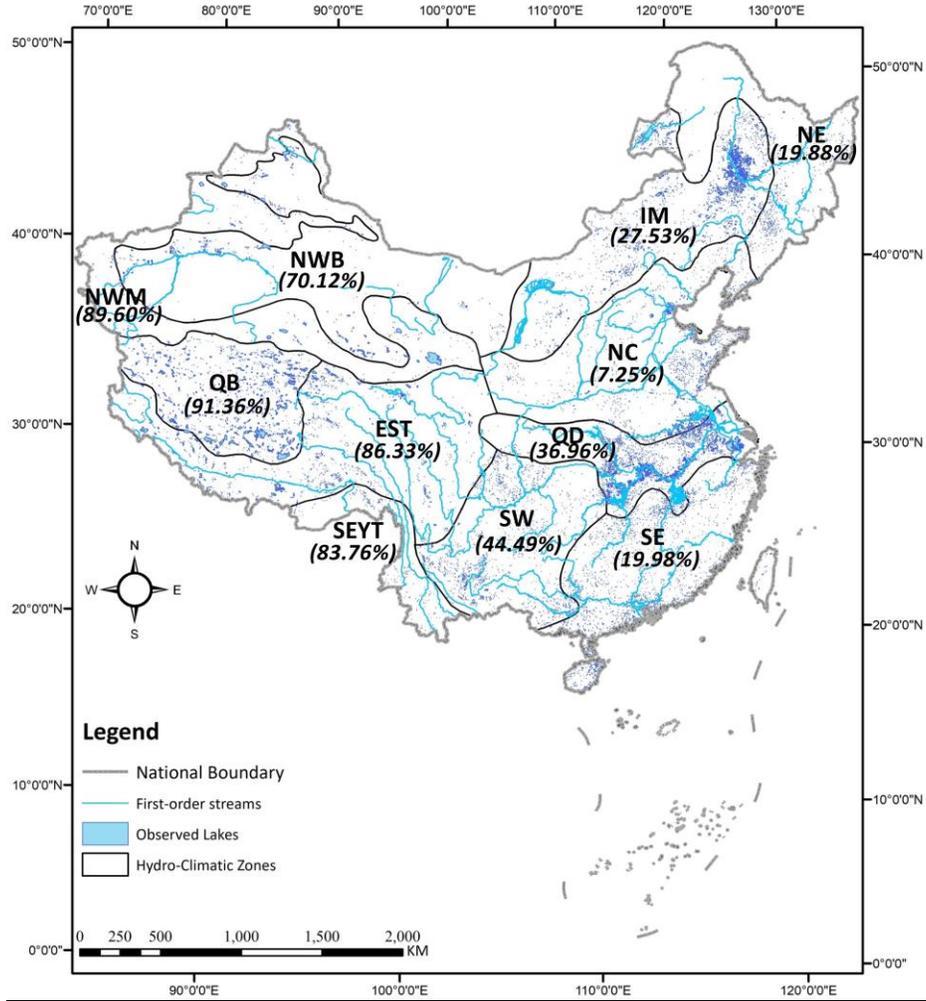

**Figure. 1.** Spatial distribution of observed (natural) lakes in China's eleven hydro-climatic zones. The conformity ratio of power law phase (*R*) is given in the bracket.

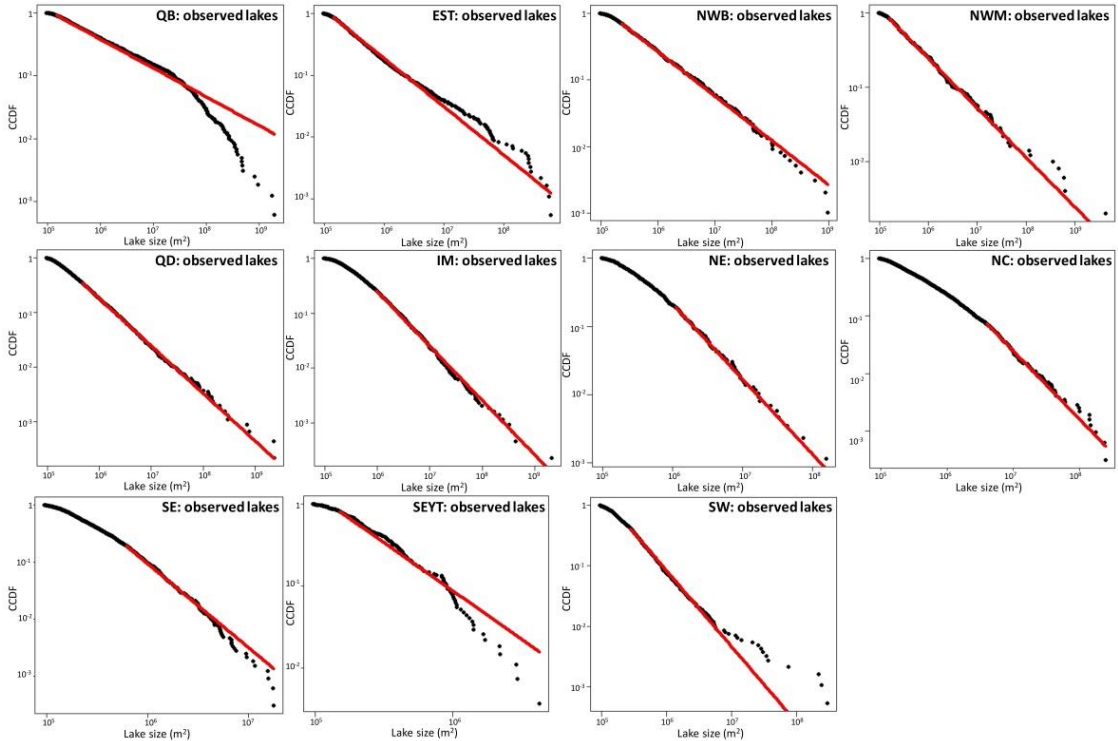

**Figure. 2**. Size distributions of observed lakes in the hydro-climatic zones (the red line indicates power law distribution).

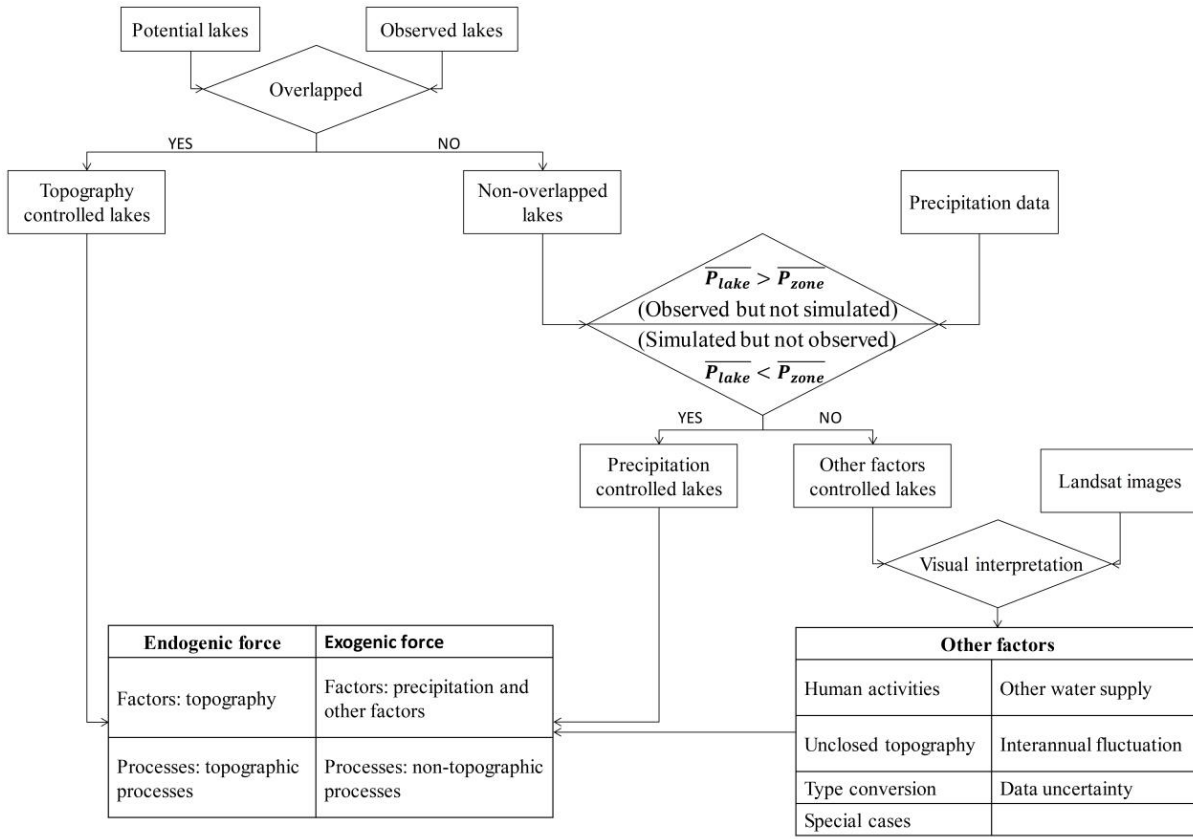

**Figure. 3**. Procedure for isolating driving factors, processes and forces of lakes.

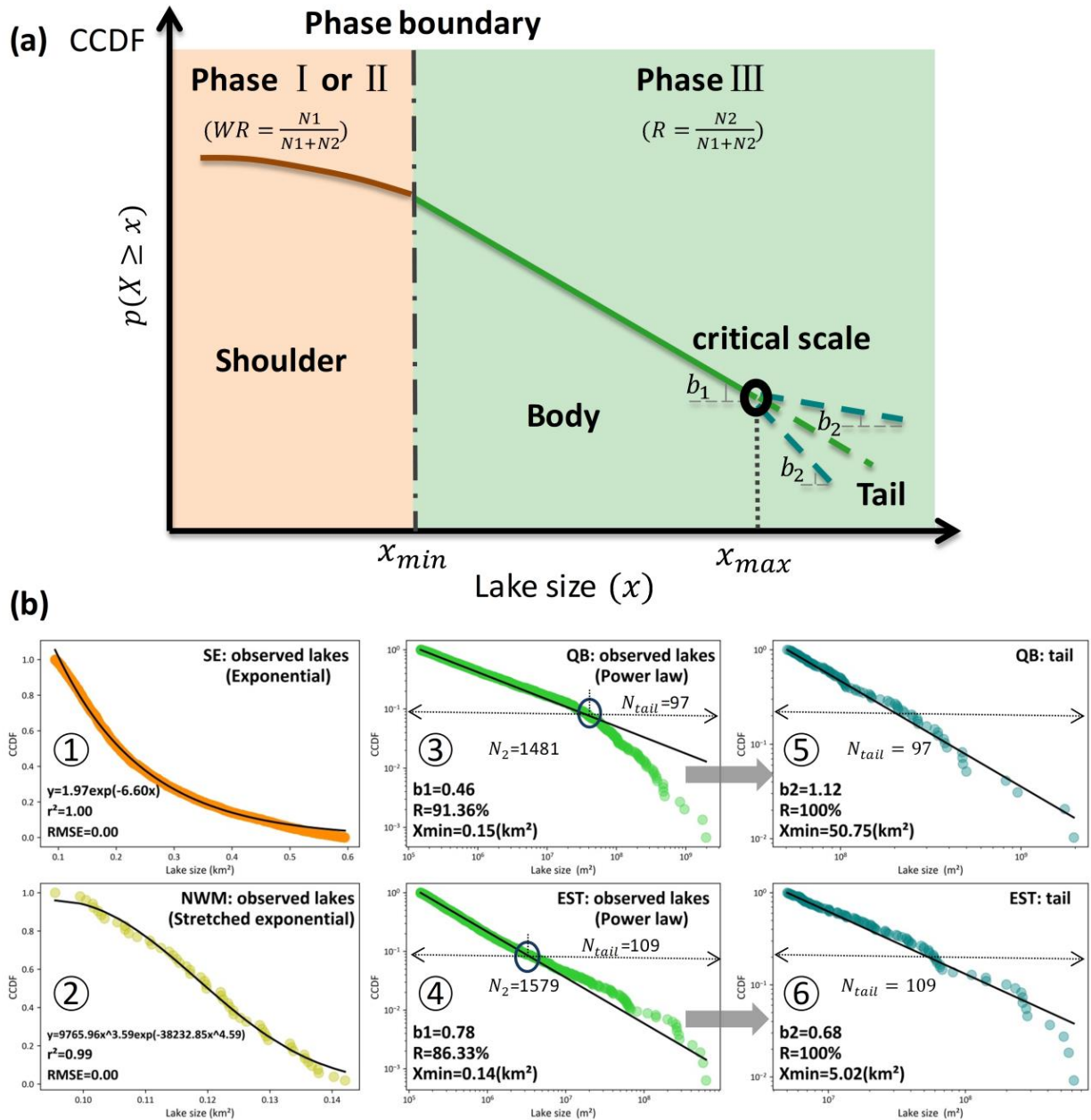

**Figure. 4.** Schematic diagram of lake size distribution along with data analysis examples. (a) the three components (shoulder, body and tail) and three phases (I: exponential; II: stretched-exponential; III: power law); (b) Data fitting results from four hydro-climatic zones. ① Exponential phase exemplified by the lakes in the shoulder component of the Southeast Zone (SE) zone; ②Stretched-exponential phase exemplified by the lakes in the shoulder component of the Northwest mountainous Zone (NWM); ③ - ⑥: Comparison of the two power law distributions coexisting in the power law phase between the two Tibetan Plateau zones (Qiangtang Basin Zone, QB; and East and southwest of Tibet Zone, EST).

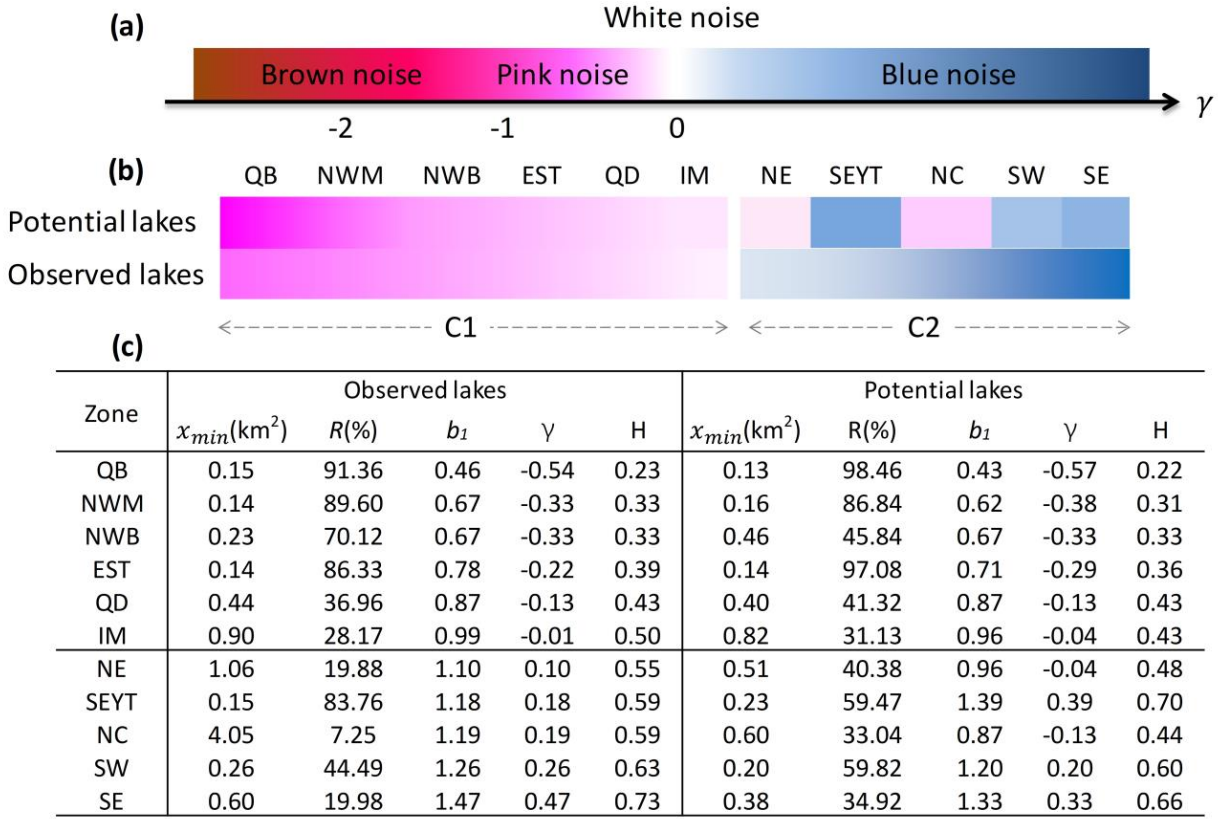

**Figure. 5.** Power spectrum analysis results of observed and potential lakes. (a) Definition of colored noise; (b) and (c): Comparison of size power spectrum characteristics between observed and potential lakes among hydro-climatic zones.

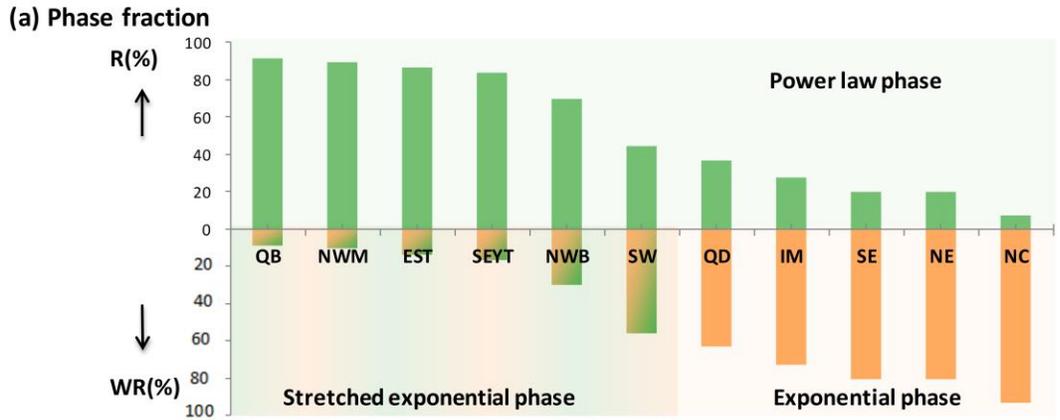

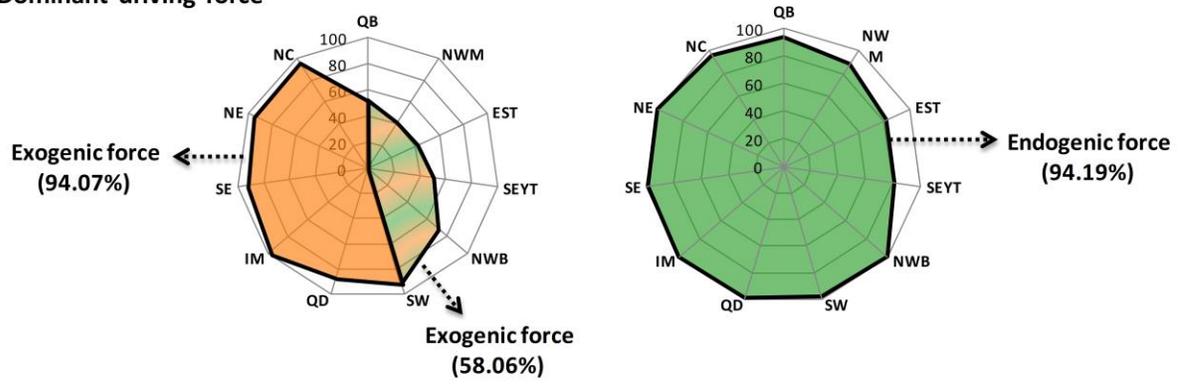

**Figure. 6.** Size distribution phases (a), its properties (b) and dominant driving forces (c). The CV and AE values in (b) are the average of the 11 hydro-climatic zones, wherein the value of AE was multiplied by 100 to be better displayed.

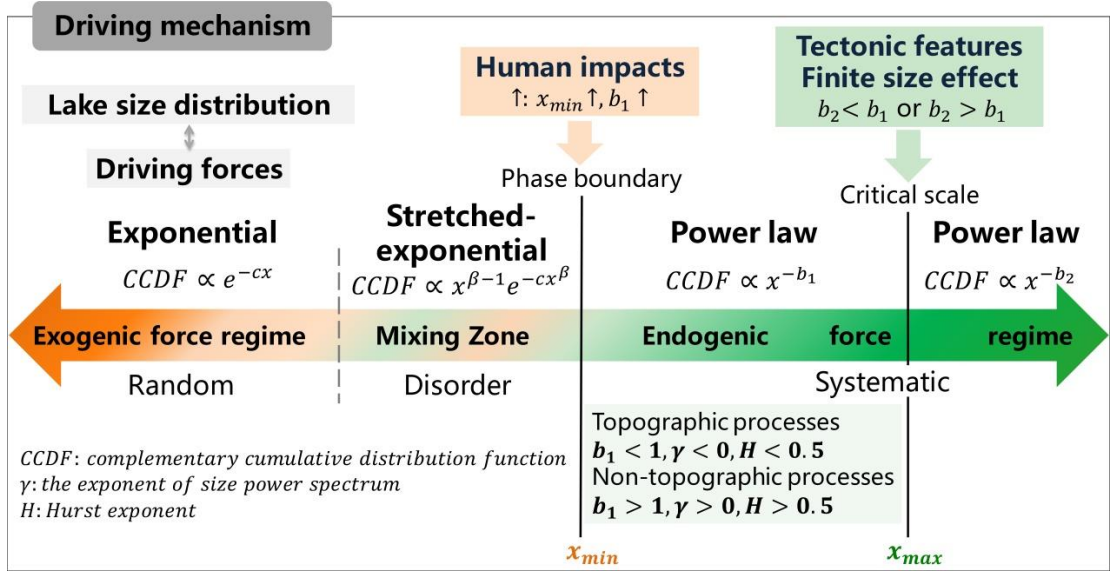

**Figure. 7.** Conceptual model of the driving mechanism underlying lake size distribution.

Table 1. Eleven hydro-climatic zones of China.

| Zone name | Abbreviation | Hydrology | Climate | Lake abundance | GDP ($10^4$ RMB/km$^2$) | HMI (High) | Rive-lake connectivity | |
|---|---|---|---|---|---|---|---|---|
| | | | | | | | Power law-distributed lakes | All the lakes |
| Southeast Zone | SE | Abundant | Tropical, subtropical | 3273 | 1897.72 | 12.70% | 1.68% | 0.82% |
| Southwest Zone | SW | Wet | Tropical, subtropical | 1870 | 899.82 | 11.82% | 1.08% | 1.50% |
| North China Zone | NC | Relatively dry | Warm temperate | 3143 | 2134.97 | 45.29% | 5.26% | 2.58% |
| Southeast of Yunnan and Tibet Zone | SEYT | Abundant | Tropical, subtropical | 271 | 96.27 | 2.15% | 0.88% | 1.11% |
| Northeast Zone | NE | Abundant, wet | Temperate, middle temperate | 860 | 225.49 | 8.85% | 2.92% | 1.40% |
| Inner Mongolia Zone | IM | Relatively dry | Middle temperate | 4356 | 376.92 | 9.13% | 1.25% | 0.85% |
| Qinba - Dabie Zone | QD | Abundant | Subtropical | 4562 | 2944.06 | 26.44% | 8.72% | 5.30% |
| Eastern and southwestern of Tibet Zone | EST | Normal | Temperate, sub frigid | 1829 | 49.69 | 0.86% | 1.71% | 1.53% |
| Northwest Mountainous Zone | NWM | Normal, relatively dry | Temperate, sub frigid, frigid | 500 | 66.15 | 2.16% | 0.89% | 0.80% |
| Northwest Basin Zone | NWB | Dry | Temperate, warm temperate | 974 | 53.74 | 1.20% | 2.93% | 2.77% |
| Qiangtang Basin Zone | QB | Relatively dry | Sub frigid, frigid | 1621 | 1.61 | 0.00% | 0.27% | 0.25% |

Table 2. Characteristics of the three components and phases of lake size distribution.

| Zone | Shoulder | | Body | | | | Tail | | | |
|---|---|---|---|---|---|---|---|---|---|---|
| | WR(%) | Type | $x_{min}$(km²) | R(%) | $b_1$ | class | Direction | $x_{max}$(km²) | R(%) | $b_2$ |
| QB | 8.64 | Stretched-exponential | 0.15 | 91.36 | 0.46 | C1 | downward | 50.75 | 100 | 1.12 |
| EST | 13.67 | | 0.14 | 86.33 | 0.78 | | upward | 5.02 | 100 | 0.68 |
| NWB | 29.88 | | 0.23 | 70.12 | 0.67 | | no deviation | | | |
| NWM | 10.40 | | 0.14 | 89.60 | 0.67 | | no deviation | | | |
| QD | 63.04 | Exponential | 0.44 | 36.96 | 0.87 | C2 | no deviation | / | / | / |
| IM | 72.47 | | 0.90 | 27.53 | 0.99 | | | | | |
| NE | 80.12 | | 1.06 | 19.88 | 1.10 | | | | | |
| NC | 92.75 | | 4.05 | 7.25 | 1.19 | | | | | |
| SE | 80.02 | | 0.60 | 19.98 | 1.47 | | downward | | | |
| SEYT | 16.24 | Stretched-exponential | 0.15 | 83.76 | 1.18 | | downward | | | |
| SW | 55.51 | | 0.26 | 44.49 | 1.26 | | upward | | | |

**Table 3.** Properties (coefficient of variation (CV) and average entropy (AE)) and dominant driving force of different phases in each hydro-climatic zone. The value of AE was multiplied by 100 to be better displayed.

| Stretched exponential | | | | Power law | | | |
|---|---|---|---|---|---|---|---|
| Zone | CV | AE | Exogenic (%) | Zone | CV | AE | Endogenic (%) |
| SW | 0.27 | 7.71 | 90.98 | SE | 1.15 | 1.89 | 98.78 |
| SEYT | 0.11 | 14.00 | 50.54 | SW | 7.85 | 2.9 | 98.08 |
| EST | 0.10 | 14.02 | 42.25 | NC | 1.9 | 1.22 | 95.61 |
| NWM | 0.10 | 13.98 | 41.61 | SEYT | 10.7 | 3.85 | 81.06 |
| NWB | 0.22 | 8.39 | 70.80 | NE | 2.7 | 1.84 | 100.00 |
| QB | 0.11 | 13.50 | 52.15 | IM | 8.6 | 0.97 | 100.00 |
| **Exponential** | | | | | | | |
| SE | 0.51 | 4.57 | 92.57 | QD | 10.85 | 1.36 | 99.11 |
| NC | 1.16 | 3.41 | 97.00 | EST | 7.2 | 2.73 | 81.76 |
| NE | 0.64 | 3.28 | 96.03 | NWM | 12.55 | 2.44 | 88.39 |
| IM | 0.56 | 3.12 | 97.93 | NWB | 6.65 | 1.78 | 99.56 |
| QD | 0.41 | 5.17 | 86.82 | QB | 5.4 | 1.34 | 93.72 |